\documentclass[fleqn,10pt]{wlscirep}
\usepackage[utf8]{inputenc}
\usepackage[T1]{fontenc}
\usepackage{bm}
\usepackage{acronym}
\usepackage{fullpage}
\usepackage{ifthen}   %
\usepackage{graphicx,graphics}
\usepackage{sidecap}
\usepackage[]{rotating}
\usepackage{setspace}
\usepackage{amssymb,amsmath,amsthm}
\usepackage{color}
\usepackage{subfigure,wrapfig}
\usepackage{multirow}
\usepackage{pdfpages}
\usepackage{longtable} 	%
\usepackage{datetime}
\usepackage{xspace}
\usepackage{array,ragged2e}
\usepackage{physics}
\usepackage{enumitem} %
\usepackage[T1]{fontenc}
\usepackage{subfigure}
\usepackage{comment}

\usepackage{xcolor}
\usepackage{braket}
\usepackage{bm}
\usepackage{nicematrix}  %

\usepackage{ifthen}
\usepackage{xkeyval}
\usepackage{calc}

\usepackage{glossaries}

\makeglossaries

\newglossaryentry{financial derivative}
{
    name=financial derivative,
    description={A financial contract that derives its value from the performance of an underlying entity}
}

\newglossaryentry{Black--Scholes model}
{
    name=Black--Scholes model,
    description={A mathematical model for the dynamics of a financial market containing derivative investment instruments}
}

\newglossaryentry{option}
{
    name=option,
    description={A financial contract that gives
the holder the right, but not the obligation, to buy or sell an underlying asset at an agreed-upon price and timeframe}
}

\newglossaryentry{bump-and-reprice}
{
    name=bump-and-reprice,
    description={A method to estimate the sensitivity of the price of a financial derivative with respect to an underlying parameter by evaluating the price at different values of the parameter and taking the difference}
}

\newglossaryentry{Hamilton--Jacobi--Bellman equation}
{
    name={Hamilton--Jacobi--Bellman equation},
    description={An equation that gives a necessary and sufficient condition for optimality of a control with respect to a loss function}
}

\newglossaryentry{martingale measure}
{
    name={martingale measure},
    description={A probability measure such that the conditional expectation of a random variable in a sequence given the value of a random variable prior in the sequence is equal to the value of this prior random variable on which the expectation is conditioned}
}

\newglossaryentry{Vapnik--Chervonenkis (VC) dimension}
{
    name={Vapnik--Chervonenkis (VC) dimension},
    description={A measure of the capacity of a set of functions that can be learned by a statistical binary classification algorithm, defined as the cardinality of the largest set of data points that the algorithm can always learn a perfect classifier for an arbitrary labelling}
}

\newglossaryentry{autocallable}
{
    name=autocallable,
    description={A financial product that pays the holder a high return if the value of the underlying asset passes an upside barrier}
}

\newglossaryentry{target accrual redemption forward}
{
    name={target accrual redemption forward},
    description={A financial product that allows the holder to achieve a target rate (interest rate, exchange rate, etc.) or rate range on a pre-defined schedule (for example, monthly) up to a limit on the maximum payout and under certain conditions on the extreme values of the rate observed in the market (spot rate). It achieves this goal by paying the holder a positive amount if the spot rate is higher than a target value and negative if lower, until the maximum amount of accrual has been reached or the spot rate hits certain upper and/or lower barriers}
}

\usepackage{algorithmic}
\usepackage[noresetcount,lined,boxed]{algorithm2e} %
\usepackage{gensymb}
\usepackage{nicematrix}

\usepackage{multirow}
\newcommand{\tabitem}{~~\llap{\textbullet}~~}

\usepackage{amsfonts}
\usepackage{amsmath}
\usepackage{supertabular}
\usepackage{amsthm}
\usepackage{floatrow}
\usepackage{tikz}
\usepackage{float}
\usepackage{mathtools}

\usepackage{titlesec}
  \usepackage{etoolbox}
  \makeatletter
  \patchcmd{\ttlh@hang}{\parindent\z@}{\parindent\z@\leavevmode}{}{}
  \patchcmd{\ttlh@hang}{\noindent}{}{}{}
  \makeatother

\usepackage{mdframed}                 %
\usepackage{colortbl}
\usepackage{xcolor}
\definecolor{highlight}{HTML}{cdeeff} %
\definecolor{highlight}{HTML}{DFEFF9}
\mdfdefinestyle{highlightbox}{leftmargin=0pt,rightmargin=0pt,%
innerleftmargin=3pt,innerrightmargin=2pt,%
innertopmargin=4pt,innerbottommargin=4pt,%
backgroundcolor=highlight,linecolor=white}
\definecolor{algm}{HTML}{D4EFDF} %
\mdfdefinestyle{highlightalg}{leftmargin=0pt,rightmargin=0pt,%
innerleftmargin=3pt,innerrightmargin=2pt,%
innertopmargin=4pt,innerbottommargin=4pt,%
backgroundcolor=algm,linecolor=white}
\definecolor{struct}{HTML}{F2D7D5} %
\mdfdefinestyle{highlightstruct}{leftmargin=0pt,rightmargin=0pt,%
innerleftmargin=3pt,innerrightmargin=2pt,%
innertopmargin=4pt,innerbottommargin=4pt,%
backgroundcolor=struct,linecolor=white}
\usepackage[section]{placeins}

\newcommand{\bvec}{\left(\begin{array}{c}}
\newcommand{\evec}{\end{array}\right)}

\title{Quantum computing for finance}

\author[1]{Dylan Herman$^\star$}
\author[2]{Cody Googin$^\dag$}
\author[3]{Xiaoyuan Liu$^\dag$}
\author[1]{Yue Sun$^\dag$}
\author[4]{Alexey Galda}
\author[5]{\\ Ilya Safro}
\author[1]{Marco Pistoia}
\author[6]{Yuri Alexeev}
\affil[1]{JPMorgan Chase, New York, NY, USA}
\affil[2]{University of Chicago, Chicago, IL, USA}
\affil[3]{Fujitsu Research of America, Inc., Sunnyvale, CA, USA}
\affil[4]{Menten AI, San Francisco, CA, USA }
\affil[5]{University of Delaware, Newark, DE, USA}
\affil[6]{Argonne National Laboratory, Lemont, IL, USA}
\affil[*]{Corresponding author: \href{mailto:dylan.a.herman@jpmorgan.com}{dylan.a.herman@jpmorgan.com}}
\affil[$\dag$]{These authors contributed equally.}

\begin{abstract}
Quantum computers are expected to surpass the computational capabilities of classical computers and have a transformative impact on numerous industry sectors. We present a comprehensive summary of the state of the art of quantum computing for financial applications, with particular emphasis on stochastic modeling, optimization, and machine learning. This Review is aimed at physicists, so it outlines the classical techniques used by the financial industry and discusses the potential advantages and limitations of quantum techniques. Finally, we look at the challenges that physicists could help tackle.
\end{abstract}

\begin{document}

\flushbottom
\maketitle

\thispagestyle{empty}

\section*{Key points}
    \begin{itemize}
        \item Quantum algorithms for stochastic modeling, optimization, and machine learning are applicable to a variety of financial problems.
        \item Quantum Monte Carlo integration and gradient estimation can provide quadratic speedup over classical methods, but more work is required to reduce the amount of quantum resources for early fault-tolerant feasibility and achieving an actual speedup.
        \item Financial optimization problems can be continuous (convex or non-convex), discrete, or mixed, and thus quantum algorithms for these problems can be applied.
        \item The advantages and challenges of quantum machine learning for classical problems are also apparent in finance.
    \end{itemize}
\clearpage  

\section*{Introduction}
\label{sec:intro}
Financial institutions tackle a wide array of computationally challenging problems on a daily basis. These problems include forecasting, (ranging from pricing and risk estimation to identifying anomalous transactions and customer preferences) and optimization (such as portfolio selection, finding optimal trading strategies, and hedging). Thanks to the advances in mathematical finance and computational techniques, with contributions from the financial industry and the scientific community, financial institutions have adopted a diverse set of problem-solving tools using stochastic modeling \cite{stochastic_modeling}, optimization algorithms, and  machine learning models for attacking both categories. In recent years, there has been increasing interest from both the academia and the industry in exploring whether quantum computing could be used to solve classically challenging problems. Researchers and industry practitioners have developed a variety of quantum computing algorithms for each of the aforementioned problem-solving tools \cite{alexeev2021prx}. 
For example, the quantum algorithm for Monte Carlo integration has shown promise for a quadratic speedup in convergence compared to the classical counterpart; this can greatly improve  risk modeling, where Monte Carlo integration \cite{glasserman2004monte} is an indispensable tool.
In optimization and machine learning, recent studies have demonstrated the advantages of quantum algorithms for discrete optimization, dynamic programming, boosting, and clustering, showing polynomially better complexity dependence on certain dimensions of the size of the problem than the state-of-the-art classical algorithms.
In addition, a variety of heuristic approaches (which often use a classical-quantum hybrid methodology involving variational quantum circuits) have been developed and heavily investigated. Preliminary results have shown indications of improvements provided by these heuristics over classical approaches.
Since most of these algorithms have general applicability to a broad category of problems, they also have similar potential in financial applications.

Notwithstanding the notable advances in quantum algorithms, there are still many challenges in achieving end-to-end quantum advantages on problems with commercially relevant specifications.
In particular, a large amount of work needs to be done to reduce the resource requirements in certain components of the algorithms, such as quantum embedding of classical input data, readout of the quantum output, and pre- and post-processing.
Otherwise, the complexity of these components may significantly disparage the speedup obtained from other components of the algorithm.
Moreover, current quantum hardware is still in a nascent stage, existing noisy intermediate-scale quantum devices (NISQ) have low fidelity and a small number of qubits, which severely limits the size of problems that can be solved.
As a result, executing quantum circuits on hardware often requires a significant optimization effort at both the circuit-design and compilation levels, and relies on classical computing to find optimal initial conditions and for error mitigation.
These challenges hint at research opportunities in addressing the bottlenecks of end-to-end quantum applications and the design of hardware-aware quantum algorithms. 
In addition, more opportunities may be discovered in exploiting the rich structures in financial problems when searching for a quantum speedup, introducing new heuristics for computationally hard problems, and improving classical heuristic algorithms.

\begin{center}
\begin{table}[h]
\fontsize{8}{10}\selectfont
\caption{
\textbf{Quantum algorithms for financial problems.} PDE partial differential equation, SCP symmetric cone programs, MIP mixed integer programming, ML machine learning, QBLAS quantum basic linear algebra subroutine}
\label{table:fintable}
\begin{tabular}{|l|l|l|l|l|l|}
\hline
\multicolumn{1}{|c|}{\textbf{Methodology}} 
& \multicolumn{1}{c|}{\textbf{Applications}}                                       
& \multicolumn{1}{c|}{\textbf{Algorithms}}                                  
& \multicolumn{1}{c|}{\textbf{Challenges}} 
& \multicolumn{1}{c|}{\textbf{Advantages}}
& \multicolumn{1}{c|}{\textbf{Applicability}}   
\\ \hline

\multirow{2}{*}{\begin{tabular}[c]{@{}l@{}}Stochastic\\ modeling\end{tabular}}       
& \multirow{2}{*}{\begin{tabular}[c]{@{}l@{}}\tabitem Pricing\\ \tabitem Risk modeling\end{tabular}} 
& \begin{tabular}[c]{@{}l@{}}Quantum Monte Carlo\\ integration\end{tabular} 
& Model loading
& \begin{tabular}[c]{@{}l@{}}Quadratic\\ speedup\end{tabular} 
& Long term \\ \cline{3-6} 
&
& Quantum PDE solvers                                                       
& Pre- and post-processing                 
& Unclear
& Long term
\\ \hline

\multirow{3}{*}{Optimization}       
& \multirow{3}{*}{\begin{tabular}[c]{@{}l@{}}\tabitem Portfolio management\\ \tabitem Financing \\ \tabitem Resource allocation \end{tabular}} 
& \begin{tabular}[c]{@{}l@{}}Quantum-enhanced\\ SCP solvers\end{tabular} 
& QBLAS limitations
& Unclear
& Long term \\ \cline{3-6} 
&
& \begin{tabular}[c]{@{}l@{}}Quantum-enhanced\\ MIP solvers\end{tabular}
& Oracle complexity                 
& \begin{tabular}[c]{@{}l@{}}Quadratic\\ speedup\end{tabular} 
& Long term \\ \cline{3-6} 
&
& Quantum heuristics                                                 
& Parameter tuning                
& \begin{tabular}[c]{@{}l@{}}Partial\\ evidence\end{tabular}
& Near term
\\ \hline

\multirow{2}{*}{\begin{tabular}[c]{@{}l@{}}Machine\\ learning\end{tabular}}       
& \multirow{2}{*}{\begin{tabular}[c]{@{}l@{}}\tabitem Forecasting\\ \tabitem Anomaly detection \\ \tabitem Recommendations \end{tabular}} 
& \begin{tabular}[c]{@{}l@{}}Quantum-enhanced ML\end{tabular}
& Data loading
& Unclear
& Long term \\ \cline{3-6} 
&
& \begin{tabular}[c]{@{}l@{}}Quantum-native ML%
\end{tabular}                                                      
& \begin{tabular}[c]{@{}l@{}}Parameter tuning and/or\\ sampling complexity\end{tabular}             
& Unclear
& Near term
\\ \hline

\end{tabular}
\end{table}

\end{center}

In this Review, we survey the state of the art of quantum algorithms for financial applications. We discuss important financial problems that are solved using techniques from stochastic modeling, optimization, and machine learning. For each of these areas, we review the classical techniques used and discuss whether the associated quantum algorithms could be useful. We also discuss the associated implementation challenges and identify potential research directions.
Table 1 is a high-level overview of the quantum algorithms covered in this Review. We note that similar tables summarizing quantum algorithms for finance from different perspectives are also present in prior reviews\cite{egger2020quantum, orus2019quantum}. 

There is an existing line of reviews of quantum computing for finance. For example,  ref.~\cite{egger2020quantum} focuses on covering the hardware and algorithmic work done by IBM. The article in ref.~\cite{bouland2020prospects} focuses on works done by the QC Ware team. The survey in ref.~\cite{orus2019quantum}  highlights financial applications that make use of quantum annealers, while also providing a short overview of certain aspects in quantum machine learning and quantum Monte Carlo integration for finance. Ref.~\cite{pistoia2021quantum} covers a variety of quantum machine learning algorithms applicable to finance. There are also several  problem-specific surveys such as the derivative pricing \cite{gomez2022survey}, supply chain finance \cite{griffin2021quantum} and high-frequency trading \cite{ganapathy2021quantum}. In contrast to these works, this Review takes a more holistic view and discusses a wider variety of financial applications and more recent quantum algorithms.
 There has been significant progress in using quantum-inspired approaches for finance and whereas a review of this line of work is beyond the scope of the current article, we do briefly mention quantum-inspired algorithms in the context of dequantized quantum algorithms in the \emph{Machine learning} section. We also direct the reader to existing surveys on other quantum-inspired methods such as tensor networks for machine learning \cite{wang2023tensor, sengupta2022tensor}, which have been applied to finance \cite{patel2022quantuminspired}.

We assume some familiarity with quantum computing and applied mathematics, otherwise we direct the reader to the standard introductory texts in quantum computing~\cite{nielsen2002quantum, kitaev2002classical} and mathematical finance~\cite{hull2021options,wilmott2013paul}.

\section*{Stochastic modeling}
\label{sec:stochastic_modeling}
Stochastic processes are commonly used to model phenomena in physical sciences, biology, epidemiology, and finance.
In the latter, stochastic modeling is often employed to help make investment decisions, usually with the goal of maximizing returns and minimizing risks, see Ref. \cite{follmer2011stochastic} for an introduction.
Quantities that are descriptive of the market condition, including stock prices, interest rates, and their volatilities, are often modeled by stochastic processes and represented by random variables.
The evolution of such stochastic processes is governed by stochastic differential equations (SDEs), and stochastic modeling aims to solve SDEs for the expectation value of a certain random variable of interest, such as the expected payoff of a \glspl{financial derivative} at a future time, which determines the fair value of the derivative.

Although analytical solutions for SDEs are available for a few simple cases, such as the geometric Brownian motion used in the \glspl{Black--Scholes model} for a European \glspl{option}~\cite{black2019pricing}, the vast majority of financial models involve SDEs of more complex forms that can only be solved with numerical approaches.
Moreover, even in cases where the SDE can be analytically solved, the complexity of the payoff specifications of the derivative may make it difficult to find closed-form exact solutions for moments of the payoff beyond the simplest cases of European options and certain Asian options~\cite{tse2018closed}.
Such more complex payoff types include American options and barrier options.
If the desired moments of the stochastic process can be expressed as a manageable differential equation, one numerical approach is to use high-precision non-randomized methods for approximately solving differential equations. The high precision comes at the cost of a strong dependence on the dimension of the stochastic process. 
Alternatively, one may use Monte Carlo integration (MCI) to trade precision for a much weaker dependence on the dimension through the variance of the stochastic process.
In fact, the integrals computed for financial problems typically involve stochastic processes whose variances do not grow exponentially with dimension as it can in the more general case.
The duality between directly solving the partial differential equation (PDE) governing the evolution of moments and estimating it through sampling has been exemplified in the Feynman--Kac formula \cite{kac1949distributions,feynman2005principle}. 
Both approaches are widely applicable even outside of stochastic modeling, and have sparked off developments in corresponding quantum methods. %
In the following subsections, we will overview financial applications that could benefit from quantum versions of numerical methods for estimating moments and other deterministic functions of stochastic processes.

\subsection*{Quantum methods for Monte Carlo-based pricing and risk analysis}

An important type of asset-pricing problem in finance is the pricing of derivatives. A derivative is a contract that derives its value from another source (such as a collection of assets or financial benchmarks) called the `underlying', whose value is modeled by a stochastic process.
The value of a derivative is typically calculated by simulating the dynamics of the underlying and computing the payoff accordingly. At each time step, the payoff function computes the cash flow to the owner of the contract, given the current and potentially historical states of the underlying assets. The payoff is a simple function that is determined when the derivative contract is made. 
The fair-value price of the derivative is given by the expected future cash flows, or payoffs, discounted to the current date.

Classically, one could simulate the payoff stochastic process using a model and estimate the price of the derivative with sample means. This is the MCI approach \cite{glasserman2004monte}, which makes no assumptions about the underlying model and can handle high-dimensional financial problems. In accordance with Chebyshev's inequality, the number of required samples from the model to achieve an estimation-error $\epsilon$ scales as $O(\sigma^2/\epsilon^2)$, where $\sigma^2$ is the variance of the payoff process.
However, there exists a quantum algorithm, commonly known as quantum MCI (QMCI), that can produce an $\epsilon$-estimate of the price using $O(\sigma/\epsilon)$ quantum samples with a constant success probability~\cite{heinrich2002quantum, brassard2011optimal, Montanaro_2015, cornelissen2021, optimal_qmean_est}. 
Specifically, suppose with probability $p(\omega)$, the state of the underlying follows a trajectory $\omega \in \Omega$ from an initial point in time to the maturity of the contract, and additionally suppose that $f(\omega)$ is the payoff function scaled to be in $[0, 1]$, then the following operation gives a single quantum sample
\begin{align}
        \label{eqn:probability_oracle}
        Q\ket{\boldsymbol{0}}=\sum_{\omega \in \Omega} \sqrt{f(\omega)p(\omega)}\ket{\omega}\ket{1}+ \ket{\perp}, %
\end{align}
which is produced by a unitary operator $Q$, and $\ket{\perp}$ is an unnormalized orthogonal garbage state produced due to unitarity. 
The vanilla QMCI uses quantum amplitude estimation (QAE) to estimate the probability of observing the rightmost qubit in the $\ket{1}$ state.  A necessary condition for a practical quantum advantage is that $Q$ is implementable in a time comparable to the time to generate a classical sample from $p(\omega)$. 
In addition, it has been noted that the overhead from current quantum error correcting codes could prevent the quadratic speedup from being realizable \cite{Babbush_2021}.
The good news is that the payoff functions in most derivatives have simple forms such as piece-wise linear functions, and therefore can be implemented using reversible-arithmetic techniques \cite{huang2019nearterm}. Furthermore, there are variants of QAE with reduced circuit depth \cite{suzuki2020amplitude, Grinko_2021, giurgica2020low}.

To implement $Q$, a variety of techniques have been developed. 
Specifically, the Grover--Rudolph algorithm~\cite{grover2002creating} and its approximate variant~\cite{marin2021quantum} aim to address loading distributions that are efficiently numerically integrable (for example, log-concave distributions). However, it has been shown that, when numerical methods, such as classical MCI, which we are trying to avoid, are used to integrate the distribution, QMCI does not provide a quadratic speedup when using this state preparation technique \cite{Herbert_2021}. 
For loading general $L_1$ or $L_2$ normalized functions, a number of black-box state preparation techniques have been proposed~\cite{grover2000synthesis,sanders2019black,wang2021fast,Wang2022inverse,Bausch2022fastblackboxquantum}. 
These algorithms start by preparing the target state with a success probability, and then use amplitude amplification to boost that probability.
Despite the general applicability of these algorithms, their inverse dependence on the filling ratio (the relevant norm of the function divided by the corresponding norm of the bounding box of that function) of the function being loaded, makes these algorithms much less efficient when the target distribution is concentrated around a few values.
An approximate method that loads degree-$d$ polynomial approximations of functions has been proposed in Ref.~\cite{mcardle2022quantum}, which has a similar dependence on the filling ratio.
Alternatively, one could use data-driven methods, such as variational quantum algorithms \cite{Cerezo_2021_survey}, which can be trained to load the path distribution \cite{Zoufal_2019,nakaji2022approximate}. 
Non-unitary distribution loading approaches have also been proposed~\cite{rattew2021efficient,rattew2022preparing}, although it is still an open question how these methods can be integrated into QMCI.
For loading tabulated data, one has to resort to generic data loading techniques~\cite{zhang2021lowdepth,mottonen2004transformation,Araujo2021divide}, which in general scale linearly with the number of data points to be loaded.

Furthermore, rather than loading the full payoff distribution $p(\omega)$ directly, it may be easier to load the joint distribution over path increments and use arithmetic to introduce any necessary correlations \cite{Chakrabarti_2021}. Nevertheless, this approach requires a number of qubits that scales linearly with the number of time steps, and quantum arithmetic operations can be expensive to perform. Therefore, it remains an open question whether there are alternative unitary procedures for loading the payoff distribution that avoid significant arithmetic and have sublinear scaling in the number of qubits. 
For scenarios where the SDE can only be simulated approximately, that is, when using local-volatility models~\cite{Dupire94pricingwith}, the discretization in time may introduce an additional multiplicative factor of $O(1/\epsilon)$ to the overall complexity of the MCI algorithm~\cite{duffie1995efficient}. 
To address this issue, the multi-level MCI~\cite{giles2008multilevel} was proposed, which ensures that the overall algorithm still scales as single-time-step MCI up to a polylogarithmic factor under certain assumptions on the payoff function.
This approach has been quantized in ref.~\cite{An_2021} with similar guarantees and extended to other payoff functions.

There is a deterministic classical method known as quasi-MCI, which for low-dimensional problems obtains a similar error scaling as QMCI. However, this comes at the cost of an exponential dependence on the dimension. Interestingly, for reasons not fully understood, this exponential dependence does not appear in practice for some financial problems \cite[Section 5.5]{glasserman2004monte}. Thus, a similar classic quadratic speedup can be obtained in some settings.

Because of the provable speedup in query complexity that QMCI provides for the widely used MCI method, it is not surprising that the community has started to perform application-specific resource analysis for the pricing of various types of derivatives, such as European~\cite{unaryoptionpricing2021} and Asian~\cite{Stamatopoulos_2020} options, \glspl{autocallable} and \glspl{target accrual redemption forward} (TARFs)~\cite{Chakrabarti_2021}.

For some derivatives, evaluating the expected payoff may be affected by future decisions, which can significantly complicate the pricing process.
An important example of such derivatives is an American option.
An American call option gives the holder the right to buy the underlying asset at a fixed price $K$ (strike) at any time between today ($t = 0$) and the maturity date ($t = T$).
At each time point $t \in [0, T]$, the holder must decide whether to exercise the option by comparing the payoff $\max(S_t-K, 0)$ if exercised today, where $S_t$ is the price of the underlying asset at $t$, with the expected payoff if exercised later. 
American option pricing falls under the broader class of optimal stopping problems. 
The exact solution of the value of an American option requires dynamic programming, because determining whether to exercise at $t$ requires solving the sub-problem of determining the future expected payoff from exercising after $t$, called a continuation value. 
For a more detailed discussion on dynamic programming, we refer the readers to the next section.
In finance, American option pricing is typically solved approximately by using regression to predict continuation values and computing the stopping time by backtracking from the maturity point. 
This technique is called least-squares Monte Carlo~\cite{longstaff2001valuing}. The labels used for training the regression model are typically computed using MCI, and there are versions that use QMCI~\cite{doriguello_et_al}. 

Another important task often accompanying derivative pricing is the computation of sensitivities of the derivative price to model and market parameters, which is equivalent to computing gradients of the price with respect to these input parameters.
Such gradients are often known as the `Greeks'.
Greeks allow for systematically hedging against risks associated with holding a derivative contract under market movements, and hence are a vital tool in risk management.
Classically, Greeks can be computed by `\glspl{bump-and-reprice}', which combines finite-difference with MCI.
Under certain continuity conditions of the payoff function with respect to the input parameter of interest, and when the MCI is performed using common random numbers, Greeks computed using \glspl{bump-and-reprice} can attain the same mean squared error convergence in terms of the number of samples used in MCI~\cite[Chapter 7]{glasserman2004monte}.
Consequently, the number of samples required to compute $k$ Greeks with an $\epsilon$ accuracy is $O(k\sigma^2/\epsilon^2)$.
Using quantum gradient methods~\cite{gilyen2019}, ref.~\cite{stamatopoulos2021} proposed a quantum acceleration of the classical \glspl{bump-and-reprice} method for computing Greeks, achieving a quadratic reduction in the number of samples required with respect to both $k$ and $\epsilon$, resulting in a sampling complexity of $O(\sqrt{k}\sigma/\epsilon)$.

When computing Greeks, under certain smoothness conditions, one can move the differentiation operation inside the expectation and instead compute the sample average of path-wise derivatives, that is, derivatives of the random variable on a fixed realization of the stochastic process. 
A variety of derivative payoffs satisfy such conditions, and smoothing techniques can be applied to payoffs with singularities~\cite{capriotti2010fast}.
On a classical computer, path-wise derivatives can be computed analytically using automatic differentiation (AD)~\cite{giles2006smoking}, which scales with the logarithmic error dependence of arithmetic.
In particular, the adjoint mode of automatic differentiation allows for the computation of $k$ gradients of a function with a cost (in terms of the number of function evaluations) independent of $k$, which provides a more efficient way of computing Greeks when the number of Greeks is large.
On a quantum computer, AD may also be performed using reversible arithmetic.
As noted in ref.~\cite{stamatopoulos2021}, when AD is used, one could use QAE to speedup the error convergence. %
Furthermore, one could utilize neural networks combining AD with back-propagation to simultaneously learn the integral and its partial derivatives~\cite{differentialML}.
Although the advantage of using a neural network is unclear at the moment, if the neural network can learn both the price and the Greeks at once requiring only $o(\sigma^2/\epsilon^2)$ training samples, it would be superior to \glspl{bump-and-reprice}.
Unfortunately, current architectures for quantum neural networks (QNNs) pay a classical sampling cost when computing gradients, that is using the parameter-shift rule~\cite{wierichs2022general}, which makes quantum advantage unlikely.

In addition to pricing and Greeks computation, MCI is also widely used in computing other risk metrics that involve the estimation of expected quantities from a stochastic process.
Similar to derivative pricing, QMCI can be applied to these tasks leading to a potential quantum speedup.
Specifically, quantum methodologies based on QMCI have been demonstrated for the computation of value-at-risk (VaR)~\cite{Woerner_2019} and credit risk~\cite{egger2019credit}.

\subsection*{Quantum methods for differential-equation-based pricing and risk analysis}

As mentioned earlier, the expectation of certain random variables, whose evolution is described by SDEs, can be formulated as the solution to parabolic PDEs.
This connection between SDEs and PDEs allows one to study stochastic processes using deterministic methods.
One common numerical technique for solving PDEs is the finite-difference method (FDM), which approximately transforms the differential equation into a system of linear equations on discretized grid points \cite{grossmann2007numerical}.
In ref.~\cite{Miyamoto_derivatives} an FDM-based approach combined with quantum linear systems algorithms (QLSAs) was used to solve the multi-asset Black--Scholes PDE. Whereas under various assumptions about the conditioning of the linear systems and data-access model, the quantum linear system can be solved with an exponential reduction in the dependence on the dimension, a sampling cost needs to be paid to readout the solution from the quantum state, making this technique reminiscent of classical and quantum MCI.
Nevertheless, it may still be beneficial to choose PDE solving methods over MCI, as it has been shown that the former can have better convergence on certain errors caused by the discretization in time, for example when pricing barrier options~\cite{broadie1997continuity}.

For second-order linear PDEs, it is possible to transform the PDE into a form that resembles the Schr\"{o}dinger equation and use Hamiltonian simulation techniques to simulate the dynamics generated by such PDEs. 
If the resulting evolution is non-unitary, a block encoding~\cite{gilyen2019} may be necessary to convert it into a unitary evolution.
It has been shown that if all eigenvalues of the evolution matrix are known, then an exponential speedup can be achieved in simulating the Hamiltonian dynamics compared to classical methods~\cite{linden2022quantum,gonzalez2021,jin2022quantum,jin2022quantum1,jin2022quantum2}.
However, since the solution of the PDE is encoded in a quantum state, QMCI, or QAE, is still needed to obtain the desired quantity from the PDE solution, thus adding additional overhead.

Variational quantum simulation (VQS) can also be used to simulate the solution for certain PDEs as the evolution of a quantum state.
Ref.~\cite{fontanela2021quantum} showed that this could be done for the Black--Scholes PDE.
Then ref.~\cite{alghassi2021variational} extended the approach to the more general Feynman--Kac PDEs and variants, and also discussed the potential applicability to the pricing of American options (that is the \glspl{Hamilton--Jacobi--Bellman equation})  and models with stochastic volatility. 
Ref.~\cite{Kubo_2021} used VQS to simulate the trinomial-tree model \cite{Boyle1986OptionVU} for solving SDEs and proposed a methodology for computing the expectation values of random variables from the SDE.

The aforementioned  algorithms encode the solution into a quantum state such that the computational basis states denote the discretized values of the spatial variable and the amplitudes are proportional to the corresponding values of the function. Ref.~\cite{kyriienko2021solving}, however, proposed an alternative approach which uses a QNN as a parameterized ansatz for the PDE solution.
The variational parameters are updated through gradient descent to satisfy the PDE. One thing to note about this approach~\cite{kyriienko2021solving} is that the input data is encoded into a product state. This potentially allows one to implement a more expressive version of the model using classical kernel methods~\cite{Schuld2021quantumkernels}. 
However, it is believed that variational quantum models can introduce a regularization not accessible to kernel methods~\cite{jerbibeyond2022}, still making QNNs that encode the input data into a single product state potentially useful. 
Although VQS and QNNs are often less resource-intensive than QLSAs, without further experimentation it remains unclear if there is any quantum advantage from these heuristic techniques.

\section*{Optimization}
\label{sec:optimization}
In this section, we discuss the potential of using quantum algorithms to help solve various optimization problems which are prevalent in finance. Most financial optimization problems are typically highly-constrained linear or quadratic programs with continuous variables, discrete variables or both. Quantum computing provides a variety of heuristics and algorithms for tackling such problems. In the subsections that follow, we review the various categories of financial optimization problems and a multitude of quantum algorithms that are applicable to such problems; we discuss whether they could actually be useful.

\subsection*{Quantum methods for continuous optimization}
In continuous optimization, the problem decision variables are real-valued. Convex programming \cite{nesterov1998introductory} is one area of continuous optimization for which there exist a variety of efficient classical algorithms for structured problems \cite{nemirovski2001lectures}. Notable examples of structured convex problems \cite{Panik_convexanalysis} are symmetric cone programs (SCPs), such as linear programs (LPs), second-order cone programs (SOCPs), and semidefinite programs (SDPs), which frequently appear in financial applications. Financing or cash-flow management and arbitrage detection are examples of important financial linear optimization problems. In a short-term financing problem \cite[Chapter 3]{optmethodfin}, the goal is to select cash flows -- for example, assets with fixed return rates, or lines of credit -- to match periodic quotas while maximizing assets. Since the amount of cash flow or credit obligation is directly proportional to the amount held, the various problem constraints are linear.

Portfolio optimization is the process of selecting the best set of assets and their quantities from a pool of assets being considered according to some predefined objective. The objective can vary depending on the investor's preference regarding financial risk and expected return. The Modern Portfolio Theory (MPT)  \cite{markowitzmpt1952} focuses on the trade-offs between risk and return to produce what is known as an efficient portfolio, which maximizes the expected return given a certain amount of risk. This trade-off relationship is represented by a curve known as the efficient frontier. The expected return and risk of a financial portfolio can often be modeled by looking at the mean and variance, respectively, of portfolio returns. 
The problem setup for portfolio optimization can be formulated as constrained utility maximization: 
\begin{equation}
    \max_{\vec{x} \in \mathcal{F}}  \vec{x}^{\mathsf{T}}\vec{\mu}-q\vec{x}^{\mathsf{T}}\Sigma\vec{x}
\end{equation}
over some feasible set of portfolios, $\mathcal{F}$. In portfolio optimization, $\vec{x}$ is the vector of asset allocations, $\Sigma$ is the covariance or correlation matrix of the assets, $\vec{\mu}$ is the vector of expected returns, and $q$ is a scaling factor that represents risk-adverseness.  The signs of the problem variables, $\vec{x}$, can also be used to indicate long or short positions. 
The problem is typically highly constrained.
For example, a weight may be assigned to each asset class, such as stocks, bonds, futures, and options, restricting the proportion of each in the portfolio. 
Assets within each class are then allocated according to their respective risks, returns, time to maturity, and liquidity, during which additional constraints may apply. %
In this subsection, we will consider cases where the variables are continuous-valued, but $\mathcal{F}$ can be non-convex. 
The discrete case will be addressed in the following subsection. 
Since most classical and quantum algorithms for continuous optimization target convex problems, specifically SCPs, we will also mainly focus on that. 

Considering that SCPs can be solved very efficiently classically, there is a small margin for quantum speedup in practice. Still, quantum reductions have been observed in worst-case running times in terms of the problem dimension due to the existence of fast quantum algorithms for basic linear algebra subroutines (BLASs). However, a quantum BLAS (QBLAS) \cite{chakraborty2018power, gilyen2019} comes with three notable caveats that make it challenging to compare quantum and classical worst-case complexities. First, a QBLAS operates on the spectrum of matrices and thus has polynomial dependence on the conditioning of the matrix, which classically typically appears only in iterative algorithms. Second, retrieving data from a quantum device requires sampling and consequently incurs a classical sampling cost that depends poorly on the desired error. Lastly, the algorithms require access to the classical data in quantum superposition, which can only be done efficiently without quantum memory in certain cases when the input matrices are sparse.

Classical techniques for solving SCPs fall into two categories, those with better dependence on the problem size and those with better dependence on the error in the solution. The first category is mainly dominated by the matrix multiplicative weights (MMW) meta-algorithm \cite{arora2007combinatorial}. It can be shown that the complexity for general SDP solving is\cite{2020sdpvanApeldoorn} $
    \tilde{O}(mns\gamma^4 + ns\gamma^7)$,
where $m$ is the number of constraints, $n$ is the number of variables, $s$ is the sparsity, and $\gamma$ is a function of the desired error. The notation $\tilde{O}(*)$ ignores factors of the form $O(\text{polylog}(*))$ in the complexity. Currently, the best quantum version of MMW for general SDP solving, in ref.~\cite{Apeldoorn2019ImprovementsIQ}, has complexity $
    \tilde{O}(s\sqrt{m}\gamma^4+s\sqrt{n}\gamma^{5})$, in terms of number of gates and calls to an oracle that provides matrix elements.
This has the optimal dependence on $m$ and $n$ one can hope for in general. However, when fine-tuned to the SDP relaxation of the MaxCut problem, for certain graphs, the classical MMW algorithm runs in $\tilde{O}(m)$. Since MMW already depends poorly on the desired problem error and assumes sparse matrices, quantum MMW may not suffer from the caveats of  QBLASs. Although MMW has not been extensively studied for solving SCPs in finance, the scalar multiplicative weights update method \cite{arora2012multiplicative}, which is a special case of MMW, has. The scalar multiplicative weights method is applicable to online portfolio optimization \cite{hazan2015online} -- an example of online convex optimization \cite{hazan2016introduction}. Online portfolio optimization, where the algorithm adapts its portfolio selection as the market changes, has received attention from the quantum computing community \cite{lim2022}. 

The scalar multiplicative weights framework has also been used to produce sublinear algorithms for estimating Nash equilibrium of two-player zero-sum games \cite{GRIGORIADIS199553}, a type of matrix game over simplices. Quantum algorithms for Gibbs sampling have been used to produce quadratic reductions in the dimension dependence for solving zero-sum games \cite{van2019quantum, bouland2023quantum}. These algorithms have been applied to the task of estimating a \glspl{martingale measure} from market data and derivative pricing \cite{incompletemarkets2022}. Furthermore, quantum amplitude amplification has been used to provide a quadratic reduction in dimension dependence over classical sublinear algorithms for a more general class of matrix games \cite{li2021sublinear} solved with multiplicative weights and online mirror descent in a primal-dual fashion \cite{shalev2012online}, with applications to training kernel classifiers \cite{li2019sublinear}. Online mirror descent is a generalization of the scalar multiplicative weights framework.

The second category of algorithms, called the polynomial-time methods, consists of the interior-point methods (IPMs) and cutting planes. In theory, the fastest method in this category for solving general SDPs is based on cutting planes \cite{lee2015}, whereas IPMs are the fastest for LPs \cite{cohen2021} and SOCPs \cite{monteiro2000polynomial}. Refs.~\cite{kerenidis2018quantum, Kerenidis_2021} use QBLASs, for example QLSAs, to perform Newton's method, which is a common IPM subroutine. For all three conic programs, the explicit dependence on the problem dimension is superior to the fastest classical algorithms. Unfortunately, these quantum IPMs suffer from all three aforementioned caveats with QBLASs. Classical IPMs can be implemented in polynomial time with logarithmic dependence on the conditioning of the matrix, that is through classical arithmetic. The classical sampling cost that the quantum algorithm pays effectively negates the good error dependence of classical IPMs and only solves the Newton system approximately. It was shown that this results in the quantum IPM only solving an approximation of the original optimization problem \cite{kerenidis2018quantum}. Proposals have been made to use variants of the standard classical IPM that may be better candidates for quantization \cite{augustino_2021} and combat some of the issues brought up. Still, resource estimates for simple portfolio optimization problems indicate a lack of advantage with quantum IPMs \cite{dalzell2022}.

Although not a polynomial-time algorithm, the simplex method is very fast in practice for solving LPs for which quantum algorithms for various subroutines have been proposed~\cite{quantumsimplex}. Also, QLSAs have been directly applied to portfolio optimization problems in the case with linear equality constraints, where the solution can be obtained in closed form by solving a linear system~\cite{rebentrost2018quantum, yalovetzky2021nisqhhl}. Besides suffering from the general issues with QBLASs, the overall benefits of these approaches remain indeterminate due to their limited applicability.

Currently, it seems to still be an open question whether quantum computation can provide an advantage for structured convex programs in general, and not just finance. However, for well-conditioned problems with sparseness or access to sufficient quantum memory, there may be speedups realizable on practical problems. However, these problems can already be solved fast with classical computation. 

In the continuous, non-convex setting, it has been demonstrated that convex relaxations may exist for some financial problems, leading to comparable performance as global optimizers, but with significantly reduced computational cost. 
In particular, tax-aware portfolio optimization is a variant of the standard MPT problem that accounts for tax liabilities. Whereas the tax penalty term in the cost function is non-convex, numerical evidence shows that a convex relaxation to an SOCP can result in a solution quality nearly identical to that from exact solvers~\cite{moehle2021tax}.
Similarly, sparse portfolio optimization, which imposes a constraint on the number of nonzero allocations in the solution, admits an SOCP relaxation that provide fast and relatively accurate solutions for some asset classes~\cite{bertsimas2022scalable}. For quantum algorithms, there has been work demonstrating speedups for certain continuous, non-convex landscapes \cite{qtwalks} as well as developing and benchmarking quantum analogues of gradient descent \cite{Leng2023}.

\subsection*{Quantum methods for discrete optimization}

Many optimization problems in finance require that the solutions take values from a discrete set, as opposed to a continuous spectrum. 
Examples include some of the most commonly seen optimization problems in finance, such as portfolio optimization also discussed in the previous subsection.
In many portfolio optimization problems, the optimal solution from a convex optimization may suggest the allocation of fractional units of an asset, whereas market constraints often require that the positions on these assets are integers or multiples of a fixed increment.
These financial use cases call for discrete optimization methods or integer programming, in which the variables to optimize are restricted to integers, and more generally, mixed integer programming (MIP), in which some of the variables are integers. For the unfamiliar reader, an introduction to discrete optimization can be found in Ref.~\cite{wolsey1999integer}.

One approximate approach for solving MIP problems is to convert the MIP into a continuous optimization problem by relaxing the integer constraints and then round the solution to the nearest feasible values allowed by the integer constraints.
Specifically, for a mixed integer linear programming problem, one may apply linear programming relaxation to remove all integer constraints.
In the more general case, convex hull relaxation may be used in which the feasible set of solutions is replaced by the minimal convex set that contains the feasible set.
In portfolio optimization problems, where typical values of the asset positions are much larger than the minimum allowed increment, approximated optimal solutions through relaxation are often acceptable with minor modifications.
This is usually the case with portfolios of stocks, where the minimum holding size is one share and the holding sizes are usually at least two orders of magnitude larger.
However, the fixed-income and derivatives markets usually require a much larger unit trading size, which makes the approximate solutions from the relaxed problem of unsatisfactory quality.
Therefore, these problems often have to resort to discrete optimization techniques.

Branch-and-bound (B\&B) methods form a powerful class of algorithms for solving hard optimization problems, such as MIPs, and can provide either a certificate of optimality or an optimality gap \cite{morrison2016branch}. The optimality gap is the difference between  the best known solution and a known bound. B\&B is the core algorithm in the majority of commercial solvers and is thus commonly adopted by financial institutions.
At a high-level, a B\&B algorithm constructs a tree of sub-problems, for example, convex relaxations in the case of MIP, whose solutions represent bounds on a possible solution to the non-relaxed problem one could hope to obtain. Subroutines based on heuristics are used for searching sub-problems to solve, specifically leaves in a tree, and prune them when possible, since the tree can be exponentially large in problem size. 

Quantum walk search (QWS) \cite{apers2019unified} is a powerful framework for speeding up classical search algorithms and has been applied to important optimization algorithms used in finance. It also worth noting that the overarching framework of quantum walks enables an at most quadratic reduction in the time to simulate symmetric Markov chains \cite{apers2018quantum}, which is potentially applicable to simulating stochastic processes arising in finance. Ref.~\cite{montanaro2020_bb} developed quantum-walk-search-based techniques, which provide an almost quadratic speedup for tree search in terms of the size of the tree. However, this does not allow the integration of search heuristics or early stopping, both of which have been found to significantly boost the performance of classical algorithms on typical problem instances. Advances have been made in this direction to integrate depth-first search heuristics~\cite{ambainis2017} and, more recently, a large class of practical heuristics~\cite{Chakrabarti_bandb}. However, similar to the case of quantum MCI for stochastic modeling, it's unclear whether an actual reduction in the time-to-solution could be achieved when considering the resources required to solve the sub-problems.

Simulated annealing (SA) \cite{romeo1991theoretical} is a widely used Markov chain Monte Carlo (MCMC) method \cite{levin2017markov} for combinatorial optimization. Quantum walks have been used to quadratically reduce the spectral-gap dependence of the asymptotic convergence time to an exact solution \cite{somma2007quantum, wocjan2008, Harrow_2020}. %
However, it's still unclear whether, in general, the dependence on all parameters with quantum can be made better than or even match the dependence they have with classical MCMC \cite{montanaro2016quantum}.
Furthermore, classical SA is typically used as a heuristic and not run until the Markov chains have converged to their stationary distributions~\cite{Henderson2003, li2009hybrid}. Still, Ref. \cite{Lemieux_2020} has investigated the use of quantum SA as a heuristic as well as designed gate-efficient constructions for implementing the quantum walk operator.

A framework~\cite{durrhoyer1996GroverMin, bulger2003implementing, Gilliam2021groveradaptive} was developed based on Grover's algorithm \cite{Grovers_algo}, which is a special case of QWS that makes use of global problem information for unstructured optimization. This algorithm was subsequently generalized in Ref.~\cite{2020sdpvanApeldoorn} to allow for arbitrary priors over the search space. The unstructured search framework requires the ability to query oracles for evaluating the function to optimize and the various constraints. Although providing a quadratic speedup in query complexity, over classical unstructured search, quantum unstructured search is not as resource efficient as quantum-walk-based techniques, which rely on local, as opposed to global, information \cite{ambainis2007quantum}. In the case of discrete optimization, uniform superpositions over the unconstrained search space can be prepared with low gate complexity, and for $\mathsf{NP}$ optimization problems the oracles for checking the constraints and evaluating the cost function can be implemented with an efficient gate complexity \cite{Sanders_2020}.

Lastly, there is also the short-path algorithm \cite{hastings2018short, dalzell2022mind}, which makes use of techniques from quantum-walk literature, and takes advantage of problem-specific information to achieve speedups that are superior to Grover-based algorithms for some discrete optimization problems.

In addition to the various quantized versions of classical algorithms, there are quantum heuristic algorithms \cite{sanders2020}. These consist of quantum annealing \cite{Kadowaki_1998, farhi2000quantum, Hegade_2021} and variational quantum algorithms: the quantum approximate optimization algorithm (QAOA) \cite{farhi2014quantum, Hadfield_2019}, variational quantum eigensolver (VQE)  \cite{Peruzzo_2014, Liu_2022}, and VQS \cite{Yuan_2019, McArdle_2019, effvarte}. In general, these methods can naturally handle unconstrained binary optimization problems, but have also been applied to continuous optimization problems \cite{continousvqe_2022, fernandezlorenzo2020hybrid}. It is possible to efficiently restrict the evolution of these quantum heuristics to respect binary-variable constraints  \cite{Hadfield_2019, niroula2022constrained, qaoa_qzd, qaoa_loan_loss_2021, Drieb_Sch_n_2023}, at least when implemented on a universal digital quantum computer. 

Variational quantum algorithms usually suffer from difficulties in tuning the variational parameters and hyperparameters, which by itself can be $\mathsf{NP}$-hard~\cite{bittel2021training}. In addition, for certain initializations, the gradients with respect to the parameters can vanish even when the model is far from convergence \cite{mcclean2018barren, larocca2022diagnosing}, requiring exponentially-many samples to estimate them. Still, the convergence of VQE has been demonstrated in the over-parameterized regime \cite{xuchen2022}, even with exponentially-small gradients.

Whereas existing quantum annealing devices are large-scale and do not need to meet the error-tolerance requirements of universal devices, the overall benefit of quantum annealing is yet to be determined. 
Quantum tunneling can allow for penetrating tall, thin potential barriers~\cite{Denchev_2016}. However, it is not possible to know a priori whether the potential barriers that appear in a practical problem permit tunneling. 
For highly-constrained financial optimization problems with multiple types of variables, it is often costly to transform the problem into an unconstrained quadratic binary optimization problem, especially when the number of qubits is limited. Still, due to the availability of relatively-large quantum annealers, the community has already begun experimenting with applications, such as portfolio optimization \cite{ mugel2020hybrid, mugel2020dynamic, palmeretf2022} and crash detection \cite{Financial_crash}.

Finally, in contrast to VQE and VQS, QAOA has been observed to have additional advantages, such as  only using two parameters per layer and theoretical and numerical results showing the ability to transfer optimized parameters between problem instances \cite{Akshay_2021, basso2022, sureshbabu2023parameter}. Furthermore, in some instances, the parameter optimization can be performed efficiently classically using simulators of quantum circuits \cite{Lykov_2021, lykov_diagonal,lykov2021large, LykovGPU}. There is still a lot of work that needs to be done to better understand the performance of quantum heuristic algorithms, especially on financial optimization problems \cite{he2023alignment}. 

\subsection*{Quantum methods for dynamic programming}

In some optimization problems, the information needed to make subsequent decisions is only revealed after intermediate decisions are made.
Therefore, in such cases, decisions must be made in a sequential manner, and an optimal strategy must take into account both current and future decisions.
Solving these decision-making problems often requires dynamic programming, in which a problem is solved recursively by reducing the main problem into a series of smaller sub-problems that are easier to solve.

Dynamic programming problems \cite{cormen2009introduction} are also commonly seen in finance.
In addition to the American option pricing problem mentioned in the first section, 
another important place where dynamic programming appears in finance is in the structuring of collateralized mortgage obligations (CMOs)~\cite[Chapter 15]{optmethodfin}.
A CMO bundles a pool of mortgages and rearranges their cash flows into multiple `tranches' that are paid in sequence.
The issuer of a CMO is often interested in an optimal payment schedule (structure) of these tranches to minimize the payment obligations to the CMO owners, which needs to be solved recursively as the optimal start and end times of the $k$-th tranche depend on the optimal schedule of the preceding $k-1$ tranches.
Although, to the best of our knowledge, there have not been demonstrations of quantum solutions to the structuring of CMOs and other similar financial problems, quantum algorithms have been proposed for reducing the exponential dependence of exponential-time dynamic programming for various $\mathsf{NP}$-complete problems~\cite{ambainis_dynamic}.

\section*{Machine learning}
\label{sec:ML}
In this section, we discuss the potential of using quantum algorithms to help solve machine learning (ML) tasks that arise in various financial applications. The field of ML \cite{mohri2018foundations, hastie2009elements} has become a crucial part of various applications in the finance industry. Rich historical financial data and advances in ML make it possible, for example, to train sophisticated models to detect patterns in stock markets, find outliers and anomalies in financial transactions, automatically classify and categorize financial news, and optimize portfolios \cite{pistoia2021quantum, arxiv_quant_fin_survey}.

Quantum algorithms for ML can be further divided into methods for accelerating classical techniques, typically by applying QBLASs, and quantum-native algorithms, which attempt to harness the classical intractability of quantum simulation to build more expressive models. The former typically requires an error-corrected quantum computer, whereas, it is argued that, the later may not and is more near term \cite{Cerezo_2021_survey}. The methods for accelerating classical algorithms need to overcome the various limitations of quantum linear algebra addressed in the previous section. The most notable one being data loading, and unfortunately, most of these algorithms work in a setting with quantum memory \cite{Giovannetti_2008,aaronson2015read}. It is currently unclear if such a device can be constructed. Furthermore, a variety of these approaches only provide speedups when the data has low-rank approximations, and in this setting, the speedup is not exponential as it was once thought. However, there remain some quantum ML algorithms that do not rely on this low-rank assumption and may remain impervious to dequantization \cite{tang2022dequantizing}. Further research is needed to determine if an exponential speedup of classical ML problems is still possible with quantum computing.

Quantum-native models, such as quantum neural networks, quantum circuit Born machine and quantum kernel methods, circumvent the issues of QBLASs and can be intractable to classically simulate, however, it is currently unclear if they provide advantage on classical problems. Specifically, although these methods can be more expressive, this has only been shown to provide advantage for problems involving data generated from quantum processes \cite{huang2021power}. In addition, initial numerical evidence suggests this does not extend to classical problems \cite{Slattery2022}. Parameterized quantum methods, such as QNNs, can be challenging to train in general \cite{mcclean2018barren} and do not currently have back-propagation algorithms that are as efficient as those for classical NNs. Whereas at the moment, quantum advantage on classical problems by using quantum-native methods appears to be unlikely, there is still significant algorithmic research that needs to be done to be certain. Furthermore, as quantum hardware advances, one will eventually be able to benchmark these heuristic algorithms on real-world problems.

In the following subsections, we highlight quantum ML algorithms that could be applied to financial problems and hope to entice further research towards quantum advantage.

\subsection*{Quantum methods for regression}
Regression is the process of fitting a numeric function from the training data set. This process is often used to understand how the value changes when the attributes vary, and it is a key tool for economic forecasting. For example, regression models can be used in asset pricing~\cite{gu2020empirical, differentialML} and volatility forecasting~\cite{GHYSELS200659}.
Least-squares is one of the most widely adopted regression models, and corresponding quantum algorithms have been proposed~\cite{wiebe2012quantum, wang2017quantum, Kerenidis_2020LS}. 
Moreover, quantum annealing has been used to solve least-squares regression when formulated as quadratic unconstrained binary optimization \cite{date2020adiabatic}. 
In addition to least-squares, Gaussian process regression (GPR) is another technique with applications in finance~\cite{Han2016}, but it suffers from a slow classical runtime.
Ref.~\cite{zhao2019quantum} proposed a QLSA-based algorithm that can obtain up to an exponential speedup for GPR for certain sparse, high-rank kernels.
Another study~\cite{PhysRevA.98.032309} used a QNN with a quantum feature map that encodes classical input data into a unitary with configurable parameters.
This technique is dubbed as quantum circuit learning~\cite{PhysRevA.98.032309}, and has been numerically shown to allow for a low-depth circuit, hence implying important application opportunities in quantum algorithms for finance.

\subsection*{Quantum methods for classification}
Classification is the process of placing objects into predefined groups. This type of process is also called pattern recognition. This area of ML can be used effectively in risk management and large data processing when the group information is of particular interest, for example, in credit-worthiness identification \cite{ABDOU201689} and fraud detection \cite{Awoyemi}. There are many well-known classical classification algorithms, such as linear classification, support vector machine (SVM), nearest centroid, and neural-network-based methods. 
Quantum algorithms could be used as subroutines to speed up existing classical algorithms.
Alternatively, a quantum version of such algorithms could be developed. Both cases could potentially benefit financial applications.
Sublinear quantum algorithms for training linear and kernel-based classifiers have been proposed~\cite{Li2019SublinearQA}, which gives a quadratic improvement for training classifiers with constant margin. Ref.~\cite{rebentrost2014quantum} proposed the quantum least-squares SVM, which uses QLSA. 
Ref.~\cite{Kerenidis_2021} proposed a quantum algorithm for SOCPs that was subsequently applied to the training of the classical $\ell_1$ SVM realizing a small polynomial speedup through numerical experiments.
In addition, classical SVMs using quantum-enhanced feature spaces to construct quantum kernels have been proposed~\cite{Havl_ek_2019}. 
Quantum nearest-neighbor classification algorithms based on Euclidean distance~\cite{wiebe2014quantum} and Hamming distance~\cite{ruan2017quantum} as the metric have also been investigated. Ref.~\cite{basheer2021quantum} proposed using a quantum $k$ maxima-finding algorithm to find the $k$-nearest neighbors and use the fidelity and inner product as measures of similarity . 
Various types of QNNs have also been applied to classification tasks~\cite{farhi2018classification, killoran2019continuous, henderson2020quanvolutional}. Quantum algorithms that improve the complexity of inference or training of a classical neural network have also been developed~\cite{allcock2020quantum, kerenidis2021classical}.

\subsection*{Quantum methods for boosting}
Boosting algorithms \cite{schapire2013boosting} use queries to a weak learning algorithm, which often produces models that classify or regress marginally better than a random guess, to construct a strong learner that can achieve an arbitrarily low prediction error. 
Adaptive boosting (AdaBoost)~\cite{schapire1990strength} was the first boosting algorithm and can be viewed as an instance of the multiplicative-weights update method. One component that appears in the complexity of AdaBoost is a linear dependence on the \glspl{Vapnik--Chervonenkis (VC) dimension} of the weak learners. Ref.~\cite{pmlr-v119-arunachalam20a} demonstrated a quantization of the AdaBoost algorithm, which promises a quadratic reduction in the dependence on the VC dimension. However, this is at the cost of a significantly worse dependence on the margin over random guess. 
Ref.~\cite{quantsmoothboost} quantized the Smooth boost algorithm~\cite{smoothboost}, which guarantees a similar quadratic reduction in the VC dependence, but with better margin dependence than Ref.~\cite{pmlr-v119-arunachalam20a}. A major limitation for both approaches is that they need to have access to quantum examples, which requires preparing quantum states encoding distributions over the training data. Lastly, there is also the QBoost framework \cite{neven2012qboost}, which proposes to use quantum heuristics for discrete optimization to select a linear combination of weak classifiers and has been applied to financial-risk detection \cite{Leclercfinrisk_2022}.

One very popular form of boosting is called gradient boosting~\cite{Friedman}, of which a high-performance instantiation has been dubbed extreme gradient boosting (XGBoost)~\cite{Chen_2016}. Currently, there does not appear to be any quantization of gradient boosting. Given the obvious generality of boosting, these models have been applied to a wide variety of ML applications in finance. Particularly, boosting has been applied to forecasting, such as derivative pricing~\cite{davis2020gradient}, financial-crisis prediction~\cite{CARMONA2019304}, credit-risk assessment~\cite{CHANG2018914}, and volatility forecasting~\cite{gavrishchaka2006boosting}.

\subsection*{Quantum methods for clustering}
Clustering, or cluster analysis, is an unsupervised ML task. It explores and discovers the grouping structure of the data.  
In finance, cluster analysis can be used to develop a trading approach that helps investors build a diversified portfolio~\cite{LEON20171334}. It can also be used to analyze different stocks, such that the stocks with high correlations in returns fall into one basket~\cite{TOLA2008235}.
Classically, there is the $k$–means clustering algorithm (also known as Lloyd’s algorithm), and quantum computing can be leveraged to accelerate a single step of $k$-means~\cite{lloyd2013quantum}. Ref.~\cite{wiebe2014quantum} showed that a step for $k$–means can be performed by using a number of queries that scales as $O(M\sqrt{k}\log(k)/\epsilon)$, where $M$ is the number of data points, $N$ is the dimension of the vectors, and $\epsilon$ is the error. 
Whereas a direct classical method requires $O(kMN)$, the quantum solution is substantially better under plausible assumptions, such as that the number of queries made by the algorithm is independent of the number of features.
Ref.~\cite{kerenidis2018qmeans} proposed $q$-means, which is the quantum equivalent of a perturbed version of $k$-means called $\delta$-$k$-means. The $q$-means algorithm has a running time that depends polylogarithmically on the number of data points. 
A NISQ version of $k$-means clustering using quantum computing has been proposed~\cite{khan2019kmeans}.
Ref.~\cite{Kerenidis} discussed a quantum version of expectation maximization, a common tool used in unsupervised ML.
Another version of $k$-means clustering has been developed~\cite{PhysRevA.101.012326} based on a quantum expectation-maximization algorithm for Gaussian mixture models.

Another clustering method that has achieved great success classically is spectral clustering~\cite{ng2002spectral}. However, it suffers from a fast-growing running time of $O(N^3)$, where $N$ is the number of points in the data set. 
Ref.~\cite{daskin2017quantum} used phase estimation and QAE for spectral clustering on quantum computers and Ref.~\cite{kerenidis2021quantum} proposed another quantum algorithm to perform spectral clustering.
Ref.~\cite{apers2020quantum} developed quantum algorithms using the graph Laplacian for ML applications, including spectral $k$-means clustering.
More discussions on unsupervised quantum ML techniques can be found in Refs.~\cite{Aimeur_Clustering, Aimeur_ML}. 
A technique for using quantum annealing for combinatorial clustering was described in Ref.~\cite{Kumar_2018}. Ref.~\cite{bermejo2206variational} proposed to reduce the clustering problem to an optimization problem and then solve it via a VQE combined with non-orthogonal qubit states.

\subsection*{Quantum methods for generative learning}
Unsupervised generative learning is at the forefront of deep learning research \cite{goodfellow2016deep}. The goal of generative learning is to model the probability distribution of observed data and generate new samples accordingly. 
One of the most promising aspects of achieving potential quantum advantage lies in the sampling advantage of quantum computers, especially considering that many applications in finance require generating samples from complex distributions. 
Therefore, further investigations of generative quantum ML for finance are needed. 

The quantum circuit Born machine (QCBM)~\cite{Marcello2019} directly exploits the inherent probabilistic interpretation of quantum wave functions and represents a probability distribution using a quantum pure state instead of the thermal distribution. Numerical simulations suggest that in the task of learning the distribution, QCBM at least matches the performance of the restricted Boltzmann machine and demonstrates superior performance as the model scales~\cite{coyle2021quantum}. (A Boltzmann machine is an undirected probabilistic graphical model inspired by thermodynamics~\cite{koller2009probabilistic}.)
QCBM was also used to model copulas~\cite{zhu2021generative}. 
Ref.~\cite{kyriienko2022protocols} proposed separating the training and sampling stages, and showed numerically that probability distributions can be trained and sampled efficiently, whereas SDEs can act as differential constraints on such trainable quantum models.

Bayesian networks are probabilistic graphical models~\cite{koller2009probabilistic} representing random variables and conditional dependencies via a directed acyclic graph, where each edge corresponds to a conditional dependency, and each node corresponds to a unique random variable. 
Bayesian inference on this type of graphs has many applications, such as prediction, anomaly detection, diagnostics, reasoning, and decision-making with uncertainty. 
Whereas exact inference is \#$\mathsf{P}$-hard, a quadratic speedup in certain parameters can be obtained by using quantum techniques~\cite{Low_2014}.
Mapping this problem to a quantum Bayesian network seems plausible since quantum mechanics naturally describes a probabilistic distribution. 
Ref.~~\cite{tucci1995quantum} introduced the quantum Bayesian network as an analog to classical Bayesian networks and Ref.~\cite{borujeni2021quantum} proposed a procedure to design a quantum circuit to represent a generic discrete Bayesian network. %
Potential applications in finance include portfolio simulation~\cite{KLEPAC2017391} and decision-making modeling~\cite{Moreira_2016}.

Classically, inference is usually performed by using MCMC methods to sample from the model's equilibrium distribution (the Boltzmann distribution)~\cite[Chapter 16]{goodfellow2016deep}. 
Because of the intractability of the partition function in the general Boltzmann machine, the graph structure is typically bipartite, resulting in a restricted Boltzmann machine~\cite{hinton2002training, salakhutdinov2009deep}. Quantum Boltzmann machines have been implemented with quantum annealing~\cite{benedetti2016estimation, dixit2021training, amin2018quantum}. In addition, a gate-based variational approach using variational quantum imaginary-time evolution, has been designed~\cite{Zoufal_2021}.

Generative adversarial networks (GANs) represent a powerful tool for classical ML: a generator tries to create statistics for data that mimic those of the real data set, while a discriminator tries to discriminate between the true and fake data.
Ref.~\cite{lloyd2018quantum} introduced the notion of quantum generative adversarial networks (QGANs), where the data consists of either quantum states or classical data, and the generator and discriminator are equipped with quantum information processors. The authors showed that when the data consists of samples of measurements made on high-dimensional spaces, quantum adversarial networks may exhibit an exponential advantage over classical adversarial networks. QGAN has been used to learn and load random distribution and can facilitate financial derivative pricing~\cite{Zoufal_2019}.

\subsection*{Quantum methods for feature extraction}
Feature extraction techniques aim to preprocess raw training data to either identify important components, transform the data to a more meaningful representation space, or reduce dimension \cite{guyon2008feature}. However, sometimes this can result in modifying the data in such a way that it is not human-interpretable, which makes it difficult to use such techniques in the highly-regulated financial industry. Thus, for finance, it is important to find efficient, useful and explainable feature extraction methods. One particular simple algorithm is principal component analysis (PCA), which finds a low-dimensional representation of the data on a linear manifold. 
A variety of quantum versions of PCA have been proposed~\cite{lloyd2014quantum, Yu_2019,lin2019improved,martin2021toward}, all of which encode principal components in quantum superposition.
Similar to most techniques that use QBLASs, it is particularly difficult to obtain useful classical information from the quantum output.
However, one could feed the result to other quantum-linear-algebra-based ML models. 

Interest in topological data analysis \cite{carlsson2020topological}, specifically persistent homology, has surged in the quantum community~\cite{lloyd2016quantum}.
In a practical setting, the potential speedup over classical algorithms is believed to be at most polynomial for dense clique complexes~\cite{McArdle2022}. 
In addition, various quantum optimization heuristics could be used for combinatorial feature selection \cite{grossi2022mixed, Ferrari_Dacrema_2022, Zoufal_2023, mucke2023feature}, which means choosing a subset of the input features to use according to some measure of importance.

\subsection*{Quantum methods for reinforcement learning}
Reinforcement learning (RL) is an area of ML that considers how agents ought to take actions in an environment in order to maximize their reward \cite{sutton2018reinforcement}. 
It has been applied to many financial applications, including pricing~\cite{halperin2020qlbs} and hedging~\cite{buehler2019deep} of contingent claims, portfolio allocation~\cite{benhamou2020deep}, automated trading under market frictions~\cite{deng2016deep,zhang2020deep}, market making~\cite{spooner2018market}, asset liability management~\cite{abe2010optimizing}, and optimization of tax consequences~\cite{kolm2020modern}. There has been a line of work investigating applying quantum algorithms to RL when one has access to state or actions spaces, for example access to Markov decision processes through quantum oracles. Ref.~\cite{dong2008quantum} proposed using Grover's algorithm to amplify the probability of observing actions that result in a positive reward. In addition, the authors showed that the approach makes a good trade-off between exploration and exploitation using the probability amplitude and can speed up learning through quantum parallelism. Additionally, Ref.~\cite{Cornelissen_2018} explored ways to apply QMCI and gradient estimation to quantize the policy gradient method, given access to a Markov decision process through quantum oracles.
Ref.~\cite{paparo2014quantum} demonstrated that the computational complexity of a particular model, projective simulation, can be quadratically reduced. In this scenario, the agent only requires quantum access to an internal memory that it builds by interacting with the environment.

Cutting-edge research in classical RL focuses on the approximate setting, where deep neural networks are used to handle state and action spaces that would otherwise be intractable with tabular RL methods~\cite{sutton2018reinforcement}. The benefits from quantum algorithms -- for instance, with QNNs -- in such a scenario appear to be unclear, similarly to when variational quantum models are used for supervised or unsupervised models. However, there has been some work in this direction~\cite{chen2020variational, chen2021variational, lockwood2020reinforcement, jerbi2021quantum, jerbi2021parametrized, crawford2016reinforcement, cherrat2023quantum}.

\subsection*{Dequantized algorithms}
The framework for dequantizing QBLAS-based algorithms started with the breakthrough result in Ref.~\cite{Tang_2019}. Dequantization results in a classical randomized algorithm that achieves a dimension dependence that is competitive with the best quantum algorithm when provided sampling access to a certain data structure containing low-rank data. These algorithms typically have very poor error dependence. Following Ref.~\cite{Tang_2019}, a variety of sublinear classical algorithms have been developed for solving low-rank SDPs, performing PCA, clustering, and more~\cite{Chia_2020}. These quantum-inspired algorithms can potentially provide benefits when high precision is not required and could be applied to financial applications~\cite{Arrazola_2020}. Despite these algorithms refuting much of the originally claimed exponential quantum speedups for ML, they could potentially inspire useful algorithms for high-dimensional financial problems.

\section*{Outlook}
\label{sec:conclusion}
Quantum computers are expected to surpass the computational capabilities of classical computers and provide a speed-up for real-world applications. As outlined in this Review, the financial industry deals with a variety of computationally challenging problems for which many applicable quantum algorithms exist. 
Promising provable quantum advantages have been discovered in certain black-box settings, such as MCI. 
However, it is still unclear whether any of these proposed approaches for stochastic modeling, optimization and ML can be turned into an end-to-end quantum advantage on practical problems.
In finance, computational time and accuracy can often directly translate to the profit and loss of the business for which the problems are being solved, insomuch that any actual wall-clock speedup and associated model performance improvement that new forms of computing could bring can have a tremendous impact on the financial industry. 
For example, fast and accurate evaluation of the risk metrics in derivatives trading is crucial in effectively hedging the risks especially under volatile market conditions. Fraud detection is apparently another time-sensitive use case in finance, as early and accurate detection of fraudulent activities can avoid potentially significant monetary loss and reputational damage to a financial institute.
As a result, the financial industry is perfectly positioned to be an early adopter and take full advantage of quantum computing in the field of computational finance.

We close with a discussion on existing quantum hardware and architecture challenges that, if solved, could benefit quantum algorithms for financial applications. As mentioned earlier, state preparation, specifically the loading of quantum states encoding classical probability distributions, is a common task required in quantum algorithms for stochastic modeling. 
One potential hardware feature that could reduce state preparation complexity is native multi-controlled gates, whose implementations have been proposed for various architectures including cold atoms~\cite{isenhower2011multibit}, trapped ions~\cite{Goel_2021} and superconducting qubits~\cite{roy2020programmable}.

Similarly, variational quantum algorithms may also benefit from the access to native multi-qubit gates.
In particular, when solving higher-order combinatorial optimization problems (for example, problems on hypergraphs~\cite{basso2022}), variational quantum optimization algorithms, such as QAOA, often require multi-body interactions, which can be encoded as parameterized multi-qubit entangling gates.
Therefore, having native access to such multi-qubit gates would help reduce the circuit complexity of these algorithms.
In addition to multi-qubit gates, a larger class of native two-qubit interactions also could enable one to more efficiently implement QAOA mixers for constrained problems, such as the Hamming-weight preserving XY-mixers~\cite{Hadfield_2019, niroula2022constrained}. As mentioned in the Optimization section, financial problems are typically highly constrained.
Moreover, having access to a variety of native entangling gates may also motivate the design and implementation of more efficient circuits for quantum ML~\cite{huang2021power}.

Coherent quantum arithmetic could also be beneficial to quantum algorithms for finance. 
Specifically, quantum algorithms for stochastic modeling and realizations of oracles for quantum optimization algorithms will most likely require a significant amount of reversible arithmetic. Consequently, qubit-count requirements for these algorithms could potentially be reduced with the development of efficient quantum floating-point arithmetic~\cite{haener2018quantum}.

Another hardware feature that could drastically improve the feasibility for quantum algorithms for finance is quantum memory. Most of the algorithms for continuous optimization and PDE solving can work with a classical-write, quantum-read memory~\cite{Apeldoorn2019ImprovementsIQ}.
However, it has been recently highlighted that existing quantum memory technologies have fundamental limitations that make the realization of low-cost and scalable quantum memory without active error correction highly challenging~\cite{jaques2023qram}. 
Therefore, new quantum memory architectures are likely needed in order to overcome these limitations.

Additionally, the operation clock rate of the current quantum hardware is much slower compared to classical computers; this is also a limiting factor in the practical usability of many quantum algorithms.
For example, quantum variational algorithms for ML require a large number of high-quality circuit evaluations to estimate gradients through sampling.
Therefore, improving the gate operation speed in quantum computers would reduce the overhead in quantum algorithms, hence potentially bringing the advantage over classical algorithms to a wider range of problem sizes as opposed to only in the asymptotic regime.
It is commonly observed that qubits encoded in `natural atoms', such as trapped ions, produce high-fidelity gate operations, but are typically slow. In contrast, `artificial atoms' such as superconducting qubits are known to be fast, but low fidelity. Combining these aspects is critical to make effective use of near-term quantum devices for financial applications.
Moreover, current quantum error correction techniques introduce a significant additional overhead that potentially negates certain quantum speedups for finance~\cite{Babbush_2021}. Thus, continued research into improving error correction and quantum architecture is also critical in realizing commercial applications of quantum computing in finance.

We hope that this Review has highlighted that in addition to challenges in hardware technologies there are still a lot of interesting quantum-algorithmic challenges to overcome to bring about a real-world quantum advantage. As quantum hardware advances, we expect to be able to benchmark more complex quantum algorithms, especially heuristic ones, on interesting problems.

\section*{Acknowledgement}
We would like to thank the support from Chicago Quantum Exchange.
Y.A. acknowledges support from the Office of Science, U.S. Department of Energy, under contract DE-AC02-06CH11357 at Argonne National Laboratory. D.H., Y.S. and M.P. appreciate the insightful discussions they had with their colleagues from the Global Technology Applied Research center at JPMorgan Chase.

\bibliography{ref}

\printglossaries

\section*{Disclaimer}
This paper was prepared for informational purposes with contributions from the Global Technology Applied Research center of JPMorgan Chase \& Co. This paper is not a product of the Research Department of JPMorgan Chase \& Co. or its affiliates. Neither JPMorgan Chase \& Co. nor any of its affiliates makes any explicit or implied representation or warranty and none of them accept any liability in connection with this paper, including, without limitation, with respect to the completeness, accuracy, or reliability of the information contained herein and the potential legal, compliance, tax, or accounting effects thereof. This document is not intended as investment research or investment advice, or as a recommendation, offer, or solicitation for the purchase or sale of any security, financial instrument, financial product or service, or to be used in any way for evaluating the merits of participating in any transaction.
 
\end{document}